\documentclass[a4paper,fleqn,usenatbib]{mnras}

\newcommand{\Msun}{M$_{\odot}$}

\newcommand{\Lsun}{L$_{\odot}$}

\usepackage{newtxtext,newtxmath}

\usepackage[T1]{fontenc}
\usepackage{ae,aecompl}

\usepackage{graphicx}	
\usepackage{amsmath}	
\usepackage{amssymb}	

\title[UVES Spectroscopy of T Chamaeleontis]{UVES Spectroscopy of T Chamaeleontis: Line Variability, Mass Accretion rate and Spectro-astrometric Analysis\thanks{Based on Observations collected with UVES at the Very Large
Telescope on Cerro Paranal (Chile), operated by the European Southern Observatory (ESO), in programs 090.C-0639(A) and 089.C-0299 (A)}}

\author[E. Cahill et al.]{
Eoin. Cahill,$^{1}$\thanks{E-mail: eoin,chaill@mu.ie}
Emma. T. Whelan,$^{1}$
Nuria Huelamo$^{2}$
and Juan Alcala$^{3}$
\\
$^{1}$Maynooth University Department of Experimental Physics, National University of Ireland Maynooth, Maynooth Co. Kildare, Ireland\\
$^{2}$Centro de Astrobiologia (INTA-CSIC) ESAC, Madrid, Spain\\
$^{3}$INAF-Osservatorio Astronomico di Capodimonte, Napoli, Italy
}

\date{Accepted XXX. Received YYY; in original form ZZZ}

\pubyear{2015}

\hypersetup{draft}
\begin{document}
\label{firstpage}
\pagerange{\pageref{firstpage}--\pageref{lastpage}}
\maketitle

\begin{abstract}
Although advances in exoplanet detection techniques have seen an increase in discoveries, observing a planet in the earliest stages of formation still remains a difficult task. Here four epochs of spectra of the transitional disk (TD) object T Cha are analysed to determine whether spectro-astrometry can be used to detect a signal from its proposed protoplanet, T~Cha~b. The unique properties of T Cha are also further constrained. H$\alpha$ and [O I]$\lambda$ 6300, the most prominent lines, were analysed using spectro-astrometry. H$\alpha$ being a direct accretion tracer is the target for the T~Cha~b detection while [O I]$\lambda$ 6300 is considered to be an indirect tracer of accretion. [O I]$\lambda$ 6300 is classified as a broad low velocity component (BC LVC). $\dot{M}_{acc}$ was derived for all epochs using new [O I]$\lambda$ 6300 LVC relationships and the H$\alpha$ line luminosity. It is shown that a wind is the likely origin of the [O I]$\lambda$ 6300 line and that the [O I]$\lambda$ 6300 line serves as a better accretion tracer than H$\alpha$ in this case. From the comparison between $\dot{M}_{acc}([OI])$ and $\dot{M}_{acc}(H\alpha)$ it is concluded that T Cha is not an intrinsically weak accretor but rather that a significant proportion of the H$\alpha$ emission tracing accretion is obscured. T~Cha~b is not detected in the spectro-astrometric analysis yet a detection limit of 0.5~mas is derived. The analysis in this case was hampered by spectro-astrometric artefacts and by the unique properties of T~Cha. While it seems that spectro-astrometry as a means of detecting exoplanets in TDs can be challenging it can be used to put an limit on the strength of the H$\alpha$ emission from accreting planetary companions and thus can have an important input into the planning of high angular resolution observations.

\end{abstract}




\section{Introduction}

Transitional disk (TD) objects are young stellar objects (YSOs) with accretion disks that have an inner region that is significantly lacking in dust \citep{Esp14}. These gaps were first revealed in the spectral energy distributions (SEDs) of T Tauri Stars (TTSs) \citep{Strom89} and most recently they have been investigated with high angular resolution optical, near-infrared and sub-millimeter imaging \citep{Hue18, Pohl17, Hen18, Hue15, Esp14, And12}. Several scenarios (e.g. photoevaporation, a binary system) have been suggested as a mechanism by which this clearing could occur \citep{Ale13}. However, clearing by a planet is also a strong possibility for the removal of dusty material from an accretion disk \citep{Erco17}. Therefore, YSOs with TDs present a unique opportunity to study planet formation at the earliest stages \citep{Esp14}. In order to establish the link between TDs and planet formation there has been a strong push to make the first definite detection of a planet in a TD \citep{Keppler18}. While studies have primarily focused on high angular resolution, high contrast imaging \citep{Pinte18, Sallum15, Reg18} attempts have also been made to use spectro-astrometry (SA) to search planets in TDs \citep{Mend18, Whe15}.

YSOs with TDs have been shown to still be actively accreting \citep{Manara14}. As planet formation models also predict accretion onto recently formed planets, an indirect method for identifying planetary companions in TDs would be to search for accretion signatures present as the planets form e.g. magnetospheric accretion \citep{Lov11}. As discussed in \cite{Whe15} these signatures could be detected by analysing strong accretion tracing emission lines e.g. H$\alpha$ \citep{Sallum15, Zhou14} or Pa$\beta$ \citep{Uyama17} with SA. SA has been routinely used to detect stellar companions around YSOs \citep{WheGar08} and \cite{Mend18} demonstrate that it should be possible to extend this work to planetary companions. TD objects are also often found to exhibit emission line variability \citep{Dupree12, Sch09}. This could be an advantage for any spectro-astrometric study as any reduction in the strength of the emission lines tracing the stellar accretion would benefit the detection of the planetary accretion signatures.  While it has been shown that SA should be capable of detecting planets in TDs no definitive detection has been made to date \citep{Mend18, Whe15}.

\begin{figure*}
\label{fig:fullspectrum}
\centering
	\includegraphics[width=17cm]{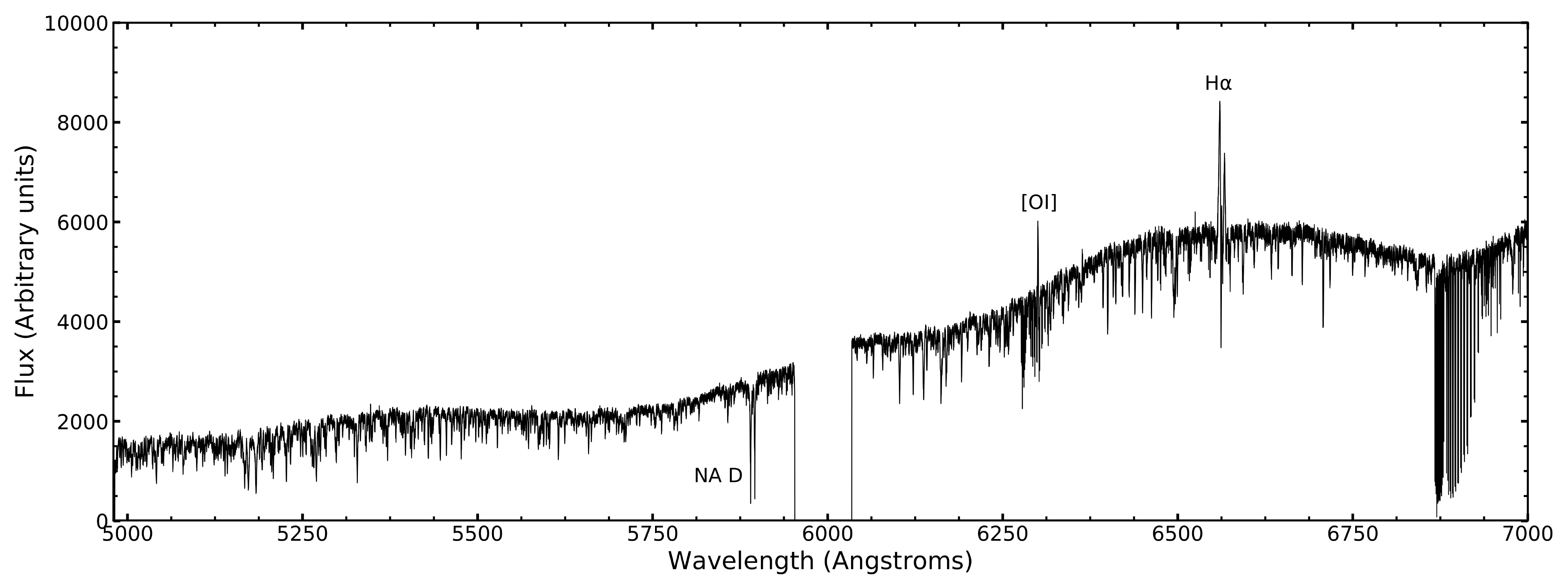}
	\caption{The full UVES spectrum of T Cha in E2 with the prominent lines marked. The spectrum is split into two parts: the lower section with a wavelength range of $\approx$ 5000 - 6000 \AA{} and the upper section with a wavelength range of $\approx$ 6000 - 7000 \AA{}}
	\end{figure*}

In this paper we report on observations of the TD object T Chamaeleontis (hereafer T Cha) taken with the ultraviolet and visual echelle spectrograph (UVES) on the European Southern Observatory's Very Large Telescope (ESO, VLT). Using sparse aperture masking observations in the infrared, \citet{Hue11} reported the detection of a substellar companion candidate within the disk gap of T Cha. However, follow-up observations of the source suggested that the detected emission was more consistent with forward scattering from the edge of the disk (see \cite{Olo13, Che15}). New ALMA millimetre observations of the source presented by \citet{Hen18} seem to be consistent with the presence of a planetary companion (or several planets) within the disk gap. The primary goal of the work presented here is to continue the investigations of \cite{Whe15} to determine if one can use SA of the H$\alpha$ line to detect sub-stellar companions. The motivation to use this technique is that planets may form very close to the central star and it is therefore difficult to directly detect them. 

\section{Target, observations and spectro-astrometry}

\subsection{T Chamaeleontis}
T Chamaeleontis ($\alpha$ = 11:54:46.7 -79:04:48.6) is a T Tauri type variable star with a TD \citep{Sch09, Hen18}. It is a member of the $\varepsilon$ Cha Association (d $\sim$ 110~pc). The distance measurement is taken from the GAIA DR2 release. Its age is given as 7~Myr by \citep{Tor08} and its mass, radius and spectral type are estimated to be 1.3 M$_{\odot}$, 1.8 R$_{\odot}$ and G8 respectively \citep{Sch09}\footnote{The stellar parameters will not change with the new GAIA DR2 distance, as in \cite{Sch09} a distance of 100pc was adopted}. \cite{Sch09} show that T Cha is still actively accreting and calculate $\dot{M}_{acc}$ to be $\sim$ 4 x 10$^{-9}$ M$_{\odot}$ yr$^{-1}$ using the 10~$\%$ width of the H$\alpha$ line. The inclination and position angle (PA) of the T Cha accretion disk are $\approx$ 69$^{\circ}$ $\pm$ 5$^{o}$ and 113$^{\circ}$ respectively \citep{Hue15, Pohl17}. The accretion disk has been found to be composed of an inner optically thick disk ranging from 0.13~au to 0.17~au \citep{Olo13}, and an outer dust disk with a radius $>$ 80~au \citep{Hue15}. These two disks are separated by a gap and it was first proposed that this gap housed a substellar companion (T Cha~b) by \cite{Hue11}. \cite{Hen18} present ALMA observations of T Cha which place an estimate of the gap to be from 18~au to 28~au. They conclude that the most likely explanation for their observations is that embedded planets are acting to carve out the dust gap. The conclusions of \cite{Hen18} support the results of \cite{Pohl17} who present modelling of VLT/SPHERE data. 

A further notable feature of T Cha is the extreme variability of its emission line spectrum \citep{Alc93}. \cite{Sch09} report strong variability in the main emission lines (H$\alpha$, H$\beta$ and [O I]) which correlates with variations in visual extinction of over three magnitudes. Both the shape and strength of the lines are variable with the H$\alpha$ line profile changing from pure emission to nearly photospheric absorption over a timescale of days. \cite{Sch09} argue that variable circumstellar extinction could be responsible for both the variations in the stellar continuum flux and for the corresponding changes in the emission lines. They propose that clumpy structures, containing large dust grains orbiting the star within a few tenths of an au, would episodically obscure the star and, eventually, part of the inner circumstellar zone. The observed radial velocity changes in the star reported by \cite{Sch09} would support this scenario. 

T Cha was chosen as a good candidate for this spectro-astrometric study due to the strong possibility that its inner disk contains a planetary companion orbiting at $>$ 20~au. As the clumpy structures proposed to be responsible for the line variability orbit closer to the star than the forming planet, any H$\alpha$ emission from a forming planet should be easier to detect during a period of obscuration of the inner circumstellar zone. 

\subsection{UVES observations}

\begin{table}
\caption{UVES observations of T Cha. E1 and E3 consist of 5 observations per slit PA. E2 and E4 consist of 8 and 16 observations per slit PA. The observation time for each epoch is 300, 600, 1000 and 600 seconds per observation respectively. The seeing is the average for each epoch and is corrected for airmass.}
	\centering
\begin{tabular}[b]{ c c c c c c }
	\hline
	Epoch & Slit PA ($^o$)&  Date &  Time (UT) & Seeing (") \\ 
	\hline
	1 & 67  & 2012-05-02  & 03:10 & 1.06  \\ 

	1 & 247  & 2012-05-02 & 03:48 & 0.89 \\

	1 & 157  & 2012-05-02 & 04:25 & 0.77 \\ 

	1 & 337  & 2012-05-02 & 05:01 & 0.74 \\

	2 & 0  & 2014-02-28 & 07:10 & 1.28 \\  

	2 & 90  & 2014-02-28 & 08:40 & 0.96 \\ 

	3 & 0  & 2014-03-04 & 04:59 & 1.24 \\ 

	3 & 90  & 2014-03-04 & 07:20 & 1.26 \\ 

	4 &  0  & 2014-03-09 & 03:04 & 0.96 \\ 

	4 & 90  & 2014-03-09 & 04:33 & 1.18 \\ 
	\hline 

\end{tabular} 
	\label{Table:1}
\end{table}
Spectra were obtained with UVES with a spectral range of $\sim$ 500 to 700~nm and a spectral resolution of R $\sim$ 40000 \citep{Dekker00} (see Figure \ref{fig:fullspectrum}). The strategy was to obtain spectra at anti-parallel and/or perpendicular slit position angles (PAs) to check for artefacts in the spectro-astrometric signature \citep{Bra06} and to recover the PA of the planetary companion. Four epochs of data (E1 to E4) were considered to allow the effect of the T Cha's variability on the chances of detecting a planet to be investigated \citep{Whe15}. Table \ref{Table:1} gives details on the observations which make up each epoch including the slit PAs used. E1 consists of UVES observations of T Cha taken in May 2012 (089.C-0299 (A)) from the ESO archive. The datasets were initially reduced using the ESO pipeline for UVES. 

\subsection{Spectro-astrometric analysis}

Spectro-astrometry is a powerful technique by which high precision spatial information can be extracted from a seeing limited spectrum. The technique involves the measurement of the centroid of the spatial profile, as a function of wavelength, to produce a so-called position spectrum \citep{WheGar08}. The precision to which the centroid can be measured is primarily dependent on the signal to noise ratio (SNR) of the spectrum and is given by 

\begin{equation}
\sigma = \frac{FWHM}{2.355(\sqrt{N_{p}})}
\label{Equation 1}
\end{equation}

where N$_{p}$ is the number of detected photons. 
SA has been used to investigate a variety of objects from brown dwarf jets to quasars \citep{Whe14, Stern15}. \cite{Whe15} discuss the technique of SA in more detail including the issue of spectro-astrometric artefacts. Possible instrumental effects which can introduce false signals to a spectro-astrometric study were noted by \cite{Bai98b} and \cite{Tak03}, and further constrained by \cite{Bra06}. The two main effects are telescope tracking errors and unstable active optics. 

For a spectro-astrometric study it is common to observe the target with different slit PAs. Taking two observations at anti-parallel PAs allows one to confirm detected signals and thus rule out artefacts \citep{Bra06}. Any real signal will invert between the two observations. An analysis of photospheric lines can also be used as a check for artefacts. The advantage of using perpendicular slit PAs is that it also enables the PA 
of the feature under investigation (outflow or companion) to be mapped \citep{Riaz17}. A further consideration for any SA project is contamination by the continuum emission. This can be corrected for by subtracting the continuum from the emission line region under analysis or by considering that the extent of any measured offset is weighted by the ratio of the continuum to line emission \citep{WheGar08}.

\begin{figure}
\centering
	\includegraphics[width=8cm]{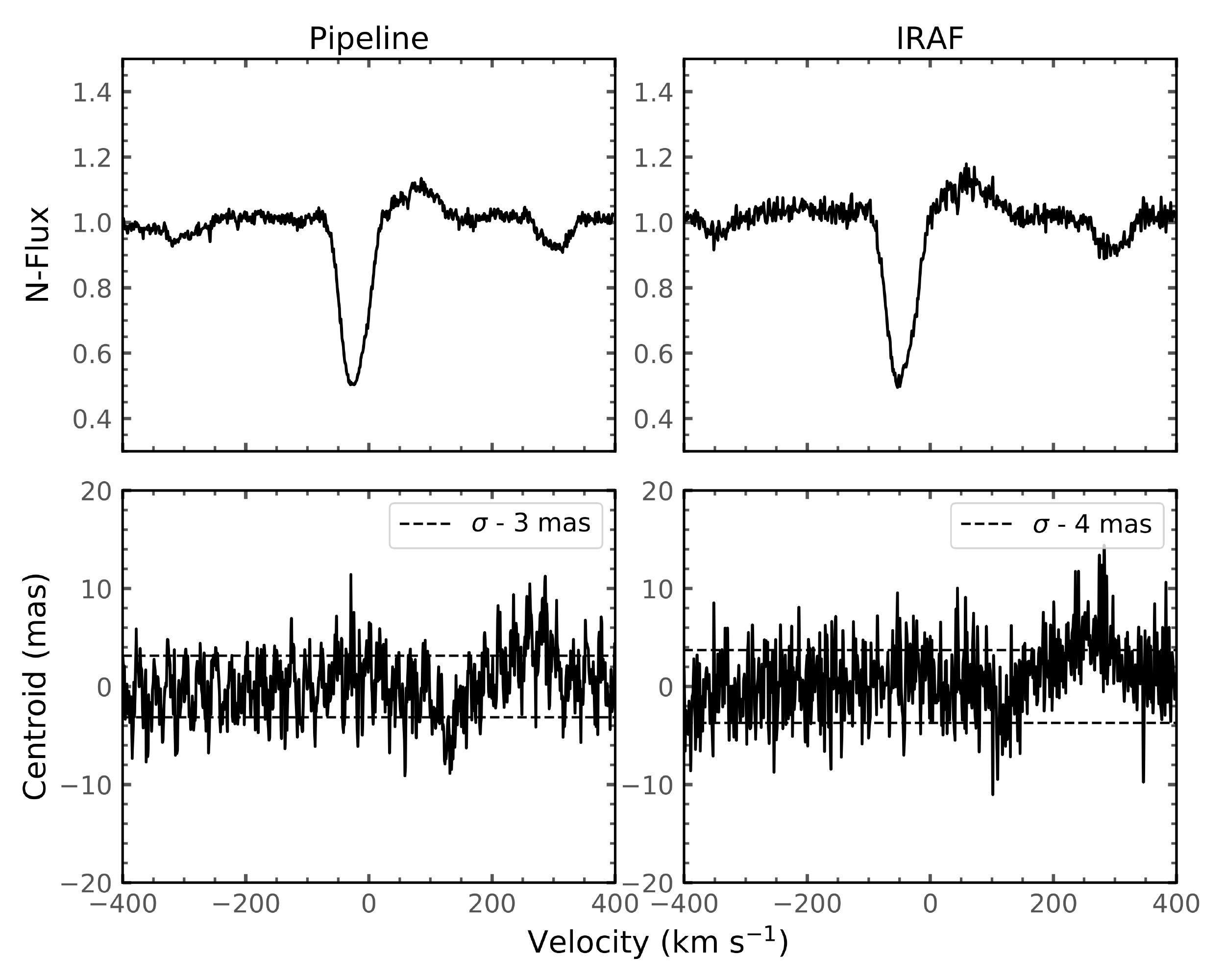}
	\caption{Left: The instrumental effect introduced by the pipeline in E1. Right: This is removed by reducing the images using IRAF routines.}
	\label{pipeline}
\end{figure}

In this work it was found that a Voigt fit provided a better fit to the wings of the spatial profile of the UVES spectra than the Gaussian fit, therefore a Voigt fit was used at all times. As discussed in \cite{Whe15} the precision in the Voigt fit can be well approximated by Equation 1. During the analysis two problems with artefacts were encountered. In \cite{Whe15} the authors discussed how X-Shooter data is affected by spatial aliasing due to the rebinning of the spectra from the physical pixel space (x,y) to the virtual pixels (wavelength, slit-scale) by the X-Shooter pipeline. The consequence of this for the SA was that an oscillating pattern on a scale of $\sim$ 10~mas was introduced to the position spectrum. In Figure \ref{pipeline} the H$\alpha$ position spectra for E1 is shown and a similar pattern is seen as was seen in the X-Shooter observations of \cite{Whe15}. It is argued that the rebinning step in the UVES pipeline is also causing a spectro-astrometric artefact due to spatial aliasing. The solution to this first artefact is not to use the pipeline reduced 2D spectra for the SA but to re-reduce the data. The spectra were therefore re-reduced using standard IRAF routines but without including a re-binning step.  

\begin{figure}
\centering
	\includegraphics[width=8cm]{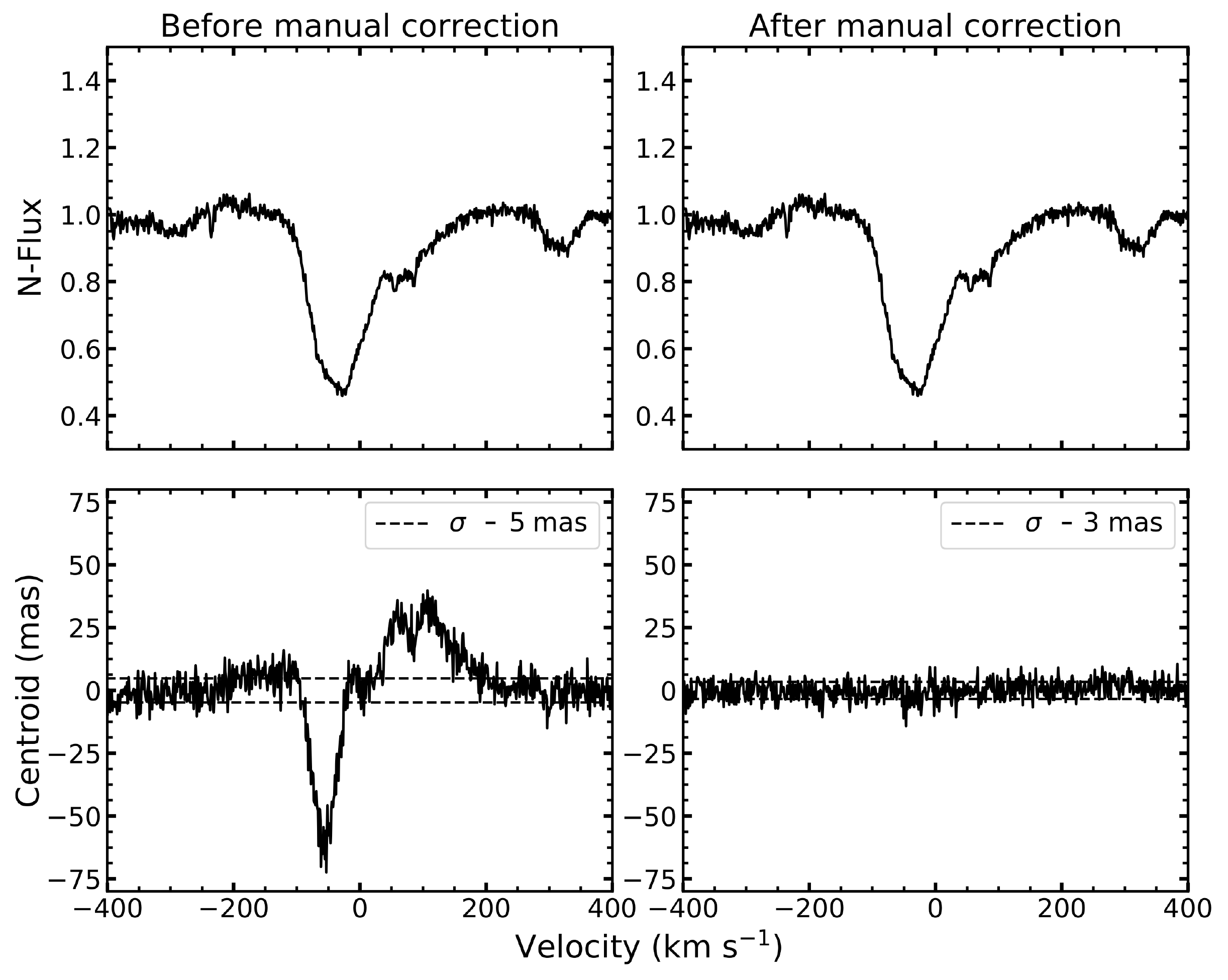}
	\caption{Comparing the unaltered median-combined centroid measurements (left) and manually re-aligned centroid and measurements (right). Both plots are taken from E3 at the zero position angle.}
	\label{fig:Artefact}
\end{figure}

\begin{figure*}
\centering
	\includegraphics[width=16cm]{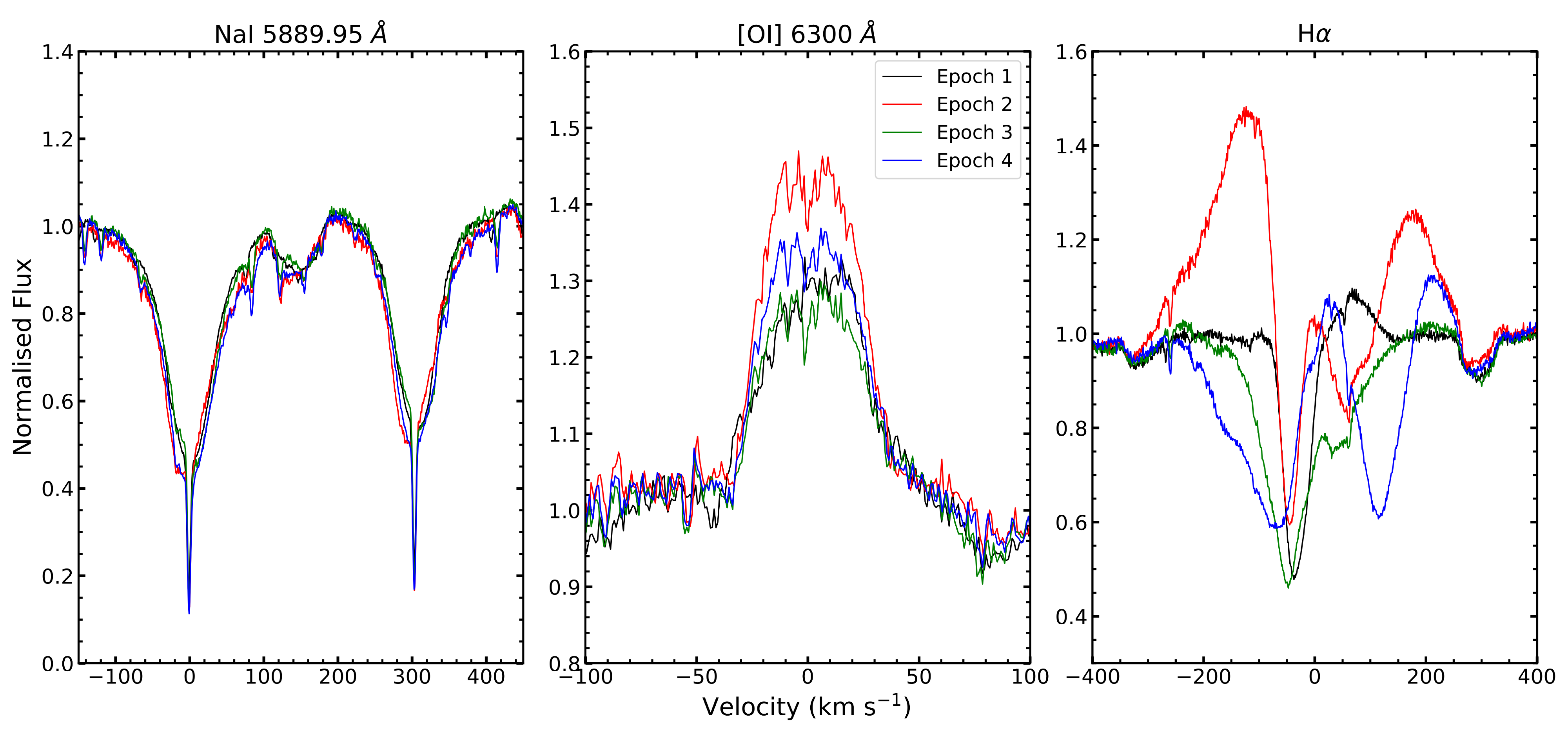}
	\caption{Comparison of line profiles across the 4 epochs, normalised to the continuum level.}
	\label{fig:Linest}
\end{figure*}

A second artefact was encountered following the median combination of some of the spectra. As shown in Table \ref{Table:1} each epoch is made up of numerous separate observations. For each epoch, each spectrum was analysed separately. Following that all the spectra making up each epoch were median combined to increase the SNR and thus the spectro-astrometric precision. For E3 strong signals were detected in the median combined data but not in the individual spectra (see Figure \ref{fig:Artefact}). Close inspection of the spectra revealed that the position of the source in the slit had drifted between observations and thus artefacts analogous to those caused by imperfect tracking were introduced when the spectra were combined. To check that this is indeed the origin of the artefact the spectra that showed the maximum drift were removed. The spectro-astrometric analysis of the median combined remaining spectra showed no signal. 

A different approach was to shift the position of the spectrum in each 2D order so that the centres of each spectra were at the same position and then to median combine the shifted spectra. As is clear from Figure \ref{fig:Artefact}, no spectro-astrometic signal was detected in the median combined shifted spectra. One of the motivations for taking several spectra rather than just one long observation was to allow for maximum consideration of the source variability. This artefact should be considered when planning future spectro-astrometric studies where variability is a factor. 

\section{Spectral analysis}

\subsection{Line emission and variability}
The full UVES spectrum for E2 is given in Figure \ref{fig:fullspectrum}. It is representative of what was detected for the other three epochs, with only the shape of the H$\alpha$ line changing (also see Figure \ref{fig:Linest}). E2 is the only epoch where any significant emission in the H$\alpha$ line is detected. This, and the absence of other emission lines seen in previous observations, indicates that T Cha is in a more quiescent state than it was for the 2002 observations of \cite{Sch09}. Despite this, the most prominent line features in the UVES spectra do agree with the results of the 2009 study. It is noted that the HeI emission line at 5876 \AA{} and the [N II]$\lambda$ 6583 forbidden line, which were detected by \cite{Sch09}, are not detected in any of the UVES observations. Also the UVES spectra did not have the necessary wavelength range to investigate H$\beta$ or the CaII triplet. \cite{Sch09} remark that the [N II]$\lambda$ 6583 line is occasionally observed in their low resolution spectra, when T Cha is in its faintest state. This is intriguing as the [N II]$\lambda$ 6583 is most often associated with high velocity jet emission yet no other signatures of a jet are detected in the T Cha spectra \citep{Whe14b}. 
\cite{Sch09} associated the He I emission with flare activity and the CaII triplet is likely tracing the accretion onto T Cha \citep{Whe14b}.

The strongest features in the UVES spectrum are the H$\alpha$, [OI]$\lambda$ 6300, and the NaD lines. In Figure \ref{fig:Linest}, these three lines are compared for the four epochs. The median profiles for each epoch are shown. The general shape of the emission lines remain the same across the observations making up the individual epochs. However, for the H$\alpha$ line changes in the relative strength of the emission and absorption features are observed. The H$\alpha$ line profiles are also shown in Figure \ref{fig:2d_results}, and the variability inherent in this line is clear, going from showing some emission in E2 to being dominated by absorption in E3 and E4. \cite{Sch09} explain these changes in the H$\alpha$ line as being due to the obscuration of T Cha by the occulting clumps and thus the suppression of the stellar continuum flux. Consequently, when the stellar continuum flux is at its lowest more of the H$\alpha$ line region is seen in emission. It is also likely that a significant part of the H$\alpha$ emission is obscured by the occulting clumps. Both accretion shocks and a stellar wind will contribute to the H$\alpha$ emission \citep{Whe14b}, however it is not clear what fraction of the H$\alpha$ emission originates close enough to the star to be obscured.

To better understand the changes in the H$\alpha$ line in the UVES dataset, the extinction was calculated for each epoch and compared to the stellar continuum flux measured in the vicinity of the H$\alpha$ line. The results are presented in Table \ref{Table:2}. In order to calculate the extinction, the observation in which H$\alpha$ is fully in absorption (E1) was selected. This provides a starting A$_{V}$ value, taken as 1.2 from \cite{Sch09}. All observations were normalized to 6349 \AA{} and the additional A$_{v}$ value required to minimize residuals between the other epochs and E1 was calculated. Comparing the H$\alpha$ line profiles with Table \ref{Table:2} it is clear that results here agree with \cite{Sch09}. The epoch in which the H$\alpha$ line is most in emission corresponds to the epoch in which the stellar continuum flux is lowest and the extinction the highest.

\begin{table*}
\caption{The extinction, stellar continuum flux, H$\alpha$ and [O I]$\lambda$ 6300 EWs, [O I]$\lambda$ 6300 radial velocity, FWHM, line luminosity and corresponding mass accretion rate for each epoch.The error in the radial velocity measurements is 2~km~s$^{-1}$. The log($\dot{M}_{acc}$) values are the mean values from the two Nisini et al. (2018) relationships. The line fluxes and consequently the line luminosities are without an extinction correction and have the continuum emission subtracted. Also shown are the values from the Covino et al. (1996) study and presented in Table 3 of Schisano et al. (2009). C1:C6 are the average values for the 6 epochs of Covino 1996 data.}
	\centering
\begin{tabular}[b]{cccccccc}
	\hline
	Epoch   &A$_{v}$  &F$_{c}$  &EW$_{H\alpha}$  &EW$_{[OI]}$ (\AA)   & V$_{[OI]}$, FWHM$_{[OI]}$   & L$_{[OI]}$    & log(M$_{acc}$)  \\ 
	&  &($\times$ 10$^{-13}$ erg/s/cm$^{2}$/\AA) &(\AA) &(\AA) &(km~s$^{-1}$) &($\times$ 10$^{-5}$ \Lsun)  & (\Msun /yr) \\
     \hline
	E1-67  &1.2 & 2.1 &0.50 &-0.43   &9.0, 54 & 3.0 $\pm$ 0.1  & -7.9  \\

	E1-157   &1.1 & 2.5 &0.44 &-0.49 &18.0, 55 & 4.1 $\pm$ 0.1   & -7.5  \\

	E1-247   &1.2 & 2.4 &0.50 &-0.48 &9.0, 57 & 4.4 $\pm$ 0.1  & -7.6  \\

	E1-337 &1.4 & 2.5 &0.41 &-0.54  &9.0, 57 & 4.3 $\pm$ 0.1& -7.6  \\

	E2-0    &1.9 & 0.5 &-1.80 &-0.58 & -3.0, 45 & 1.1 $\pm$ 0.1  & -8.5  \\
    
	E2-90  &1.9 & 0.6 &-1.2 &-0.56 &-2.0, 46 & 1.1 $\pm$ 0.1  & -8.5  \\

	E3-0  &1.5 & 0.6 &1.3 &-0.46 &-3.0, 51 & 0.8 $\pm$ 0.1  & -8.8 \\

	E3-90  &1.4 & 0.9 &1.6 &-0.43  &-2.0, 53 & 1.2 $\pm$ 0.1 & -8.5  \\

	E4-0  &1.7 & 0.8 &1.5 &-0.49 &-2.0, 47 & 1.2 $\pm$ 0.1   & -8.4   \\

	E4-90  &1.7 & 0.9 &1.3 &-0.49 &-3.0, 52 & 1.3 $\pm$ 0.1   & -8.4 \\
   
    C1:C6  & &- &-2.7 &-0.6 &1.4 $\pm$ 0.1 &- &-8.4 \\
    
     C1 &- &- &- &-0.8 &- &1.4 $\pm$ 0.1 &-8.4 
     \\
     C2  &- &-  &- &-0.9 &- &1.3 $\pm$ 0.1 &-8.4 
     \\
     C3 &- &- &- &-1.2 &- &1.5 $\pm$ 0.2 &-8.3 
     \\
     C4 &- &- &- &-0.3  &- &1.3 $\pm$ 0.1 &-8.4 
     \\
     C5 &- &- &- &-0.3 &- &1.2 $\pm$ 0.1 &-8.5 
     \\
     C6  &- &- &- &-0.2 &- &1.2 $\pm$ 0.2 &-8.5 
     \\
	\hline 
\label{Table2}
\end{tabular} 
	\label{Table:2}
\end{table*}


The forbidden [O I]$\lambda$ 6300 line is frequently found to have both a low and high velocity component (LVC, HVC) which trace a low velocity wind and a collimated jet respectively \citep{Whe04}. \cite{Sim16} present a comprehensive study of the LVC in TTSs. Using Gaussian fitting they classify a LVC component as having a centroid velocity of 30~km~s$^{-1}$ or less. They further divide the LVC into broad and narrow components (BC, NC) using the full width half maximum (FWHM) of the line to separate them. They designate the NC LVCs as those with FWHM $\leq$ 40~km~s$^{-1}$ and BC LVCs with FWHM $>$ 40~km~s$^{-1}$. Adopting this criteria, the T Cha [O I]$\lambda$ 6300 line profile with an average centroid velocity of 3~km~s$^{-1}$ and an average FWHM of 52~km~s$^{-1}$ is classified a BC LVC. This agrees with the \cite{Sch09} observation that the [O I]$\lambda$ 6300 line profile is intermediate between a LVC and a HVC. The profile remains generally constant in shape across the UVES epochs (Figure \ref{fig:Linest}) but the flux level (converted to a line luminosity in Table \ref{Table:2}) and radial velocity of the line do change, although the FWHM remains constant within errors. 

\begin{figure}
\centering
	\includegraphics[width=9cm]{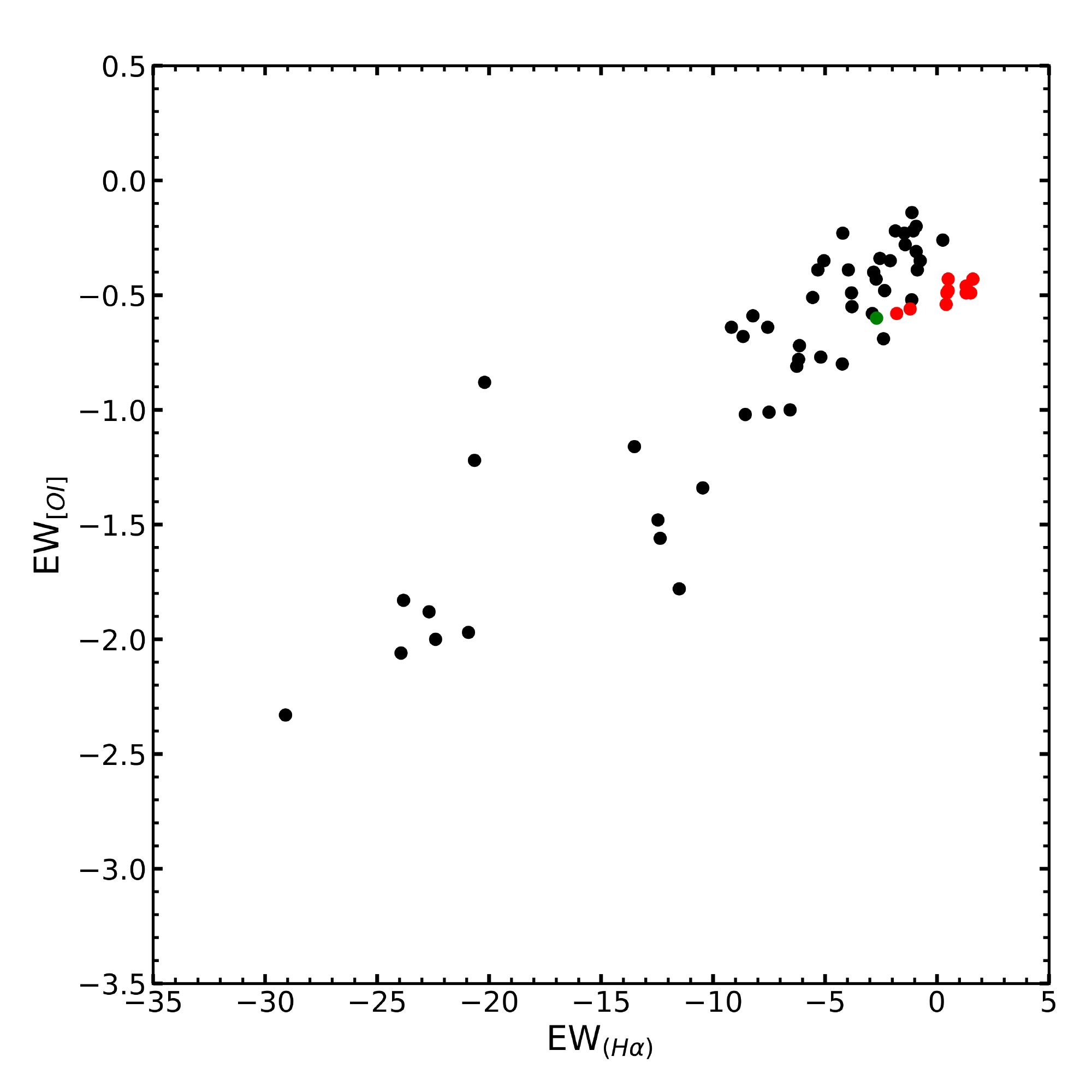}
	\caption{The correlation of the H$\alpha$ and [O I]$\lambda$ 6300 EWs. The black points are the FEROS data presented in Schisano et al. (2009). The red points are the measurements from the UVES spectra presented here and the green point is the average values from the Covino et al. (1996) dataset.}
	\label{fig:EW}
\end{figure}

\cite{Sch09} detect both a variability in the equivalent width (EW) of the [O I]$\lambda$ 6300 line and the radial velocity. They report that the line profile is generally symmetrical and centered at $\sim$ 0~km~s$^{-1}$ but that sometimes a slight red-shift of about 10~-~20~km~s$^{-1}$ is seen. They find that the variation in the EW of the [O I]$\lambda$ 6300 emission is correlated with the changes in the EW of the H$\alpha$ line. Furthermore, they investigate both the EW and luminosity of the [O I]$\lambda$ 6300 in the only simultaneous spectroscopy of T Cha available to them (the \cite{Cov96} dataset) and find that while the EW changes the luminosity is relatively constant with time. Overall, they conclude that the variability in the emission-line intensity of the [O I]$\lambda$ 6300 emission is caused by changes in the stellar continuum flux level rather than any intrinsic variability of the line-emitting region. 

In Figure \ref{fig:EW}, Figure 7 of \cite{Sch09} is replotted with the UVES EWs also shown. This comparison shows that the H$\alpha$ and [O I]$\lambda$ 6300 EWs from the UVES data follow the correlation reported by \cite{Sch09}. Here, while  L$_{[OI]}$ is stable within each epoch (Table \ref{Table:2}) a large change between E1 and E2-E4 is noted. Also note that between E2 and E4 the line luminosity does not change much. Furthermore, the epoch of highest luminosity also corresponds to the lowest values of A$_{v}$ and consequently the highest levels of continuum flux, and a significant shift in the radial velocity of the line. If the last three epochs are considered alone along with Figure \ref{fig:EW}, results support the conclusions of \cite{Sch09}. However, the changes in line luminosity and radial velocity between E1 and the other three epochs could point to some component of the [O I]$\lambda$ 6300 line being obscured by the occulting clumps, so that when the extinction drops this component becomes visible changing both the flux and velocity centroid of the line. We note that the average flux of the [O I]$\lambda$ 6300 line is 1 $\pm$ 0.1 $\times$ 10$^{-13}$ erg/cm/s$^{2}$ in E1 and decreases to 0.3 $\pm$ 0.4 $\times$ 10$^{-13}$ erg/cm/s$^{2}$ for E2-E4. An intrinsic dimming of the wind during the almost two years between E1 and E2-E4 cannot be ruled out.

\begin{figure*}
	\includegraphics[width=18cm]{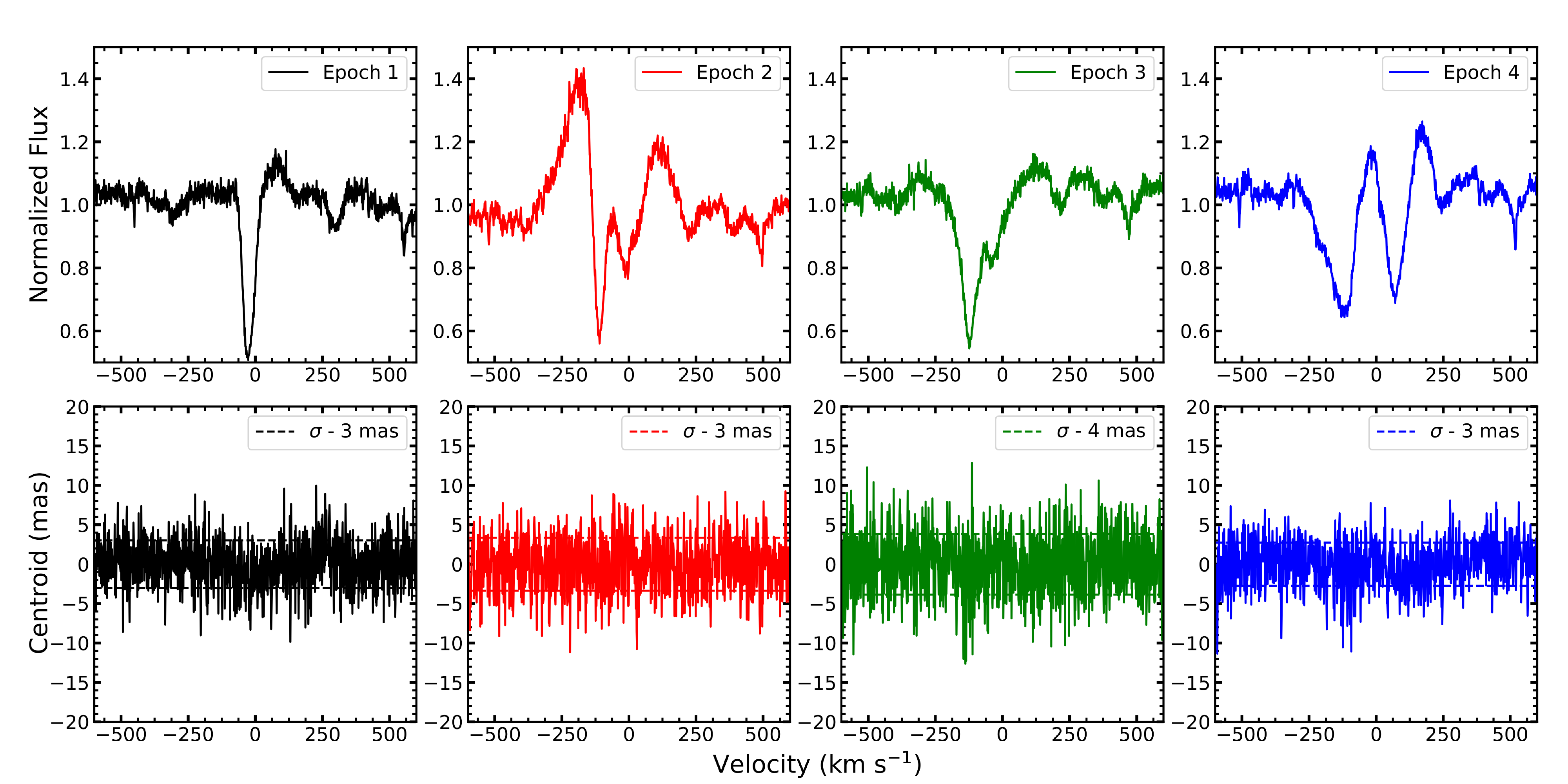}
	\caption{Spectro-astrometric analysis of the H$\alpha$ emission from T Cha. What is shown here is the median of the spectra for each of our four epochs. The spectra were checked individually and no signal was found. The accuracy achieved on average over all 4 epochs was $\sigma$ = 3.65 mas or 0.3 au at a distance of 110 pc to T Cha}
	\label{fig:2d_results}
 \end{figure*}

\subsection{Spectro-astrometric results}
\begin{figure}
	\includegraphics[width=6cm, angle=90]{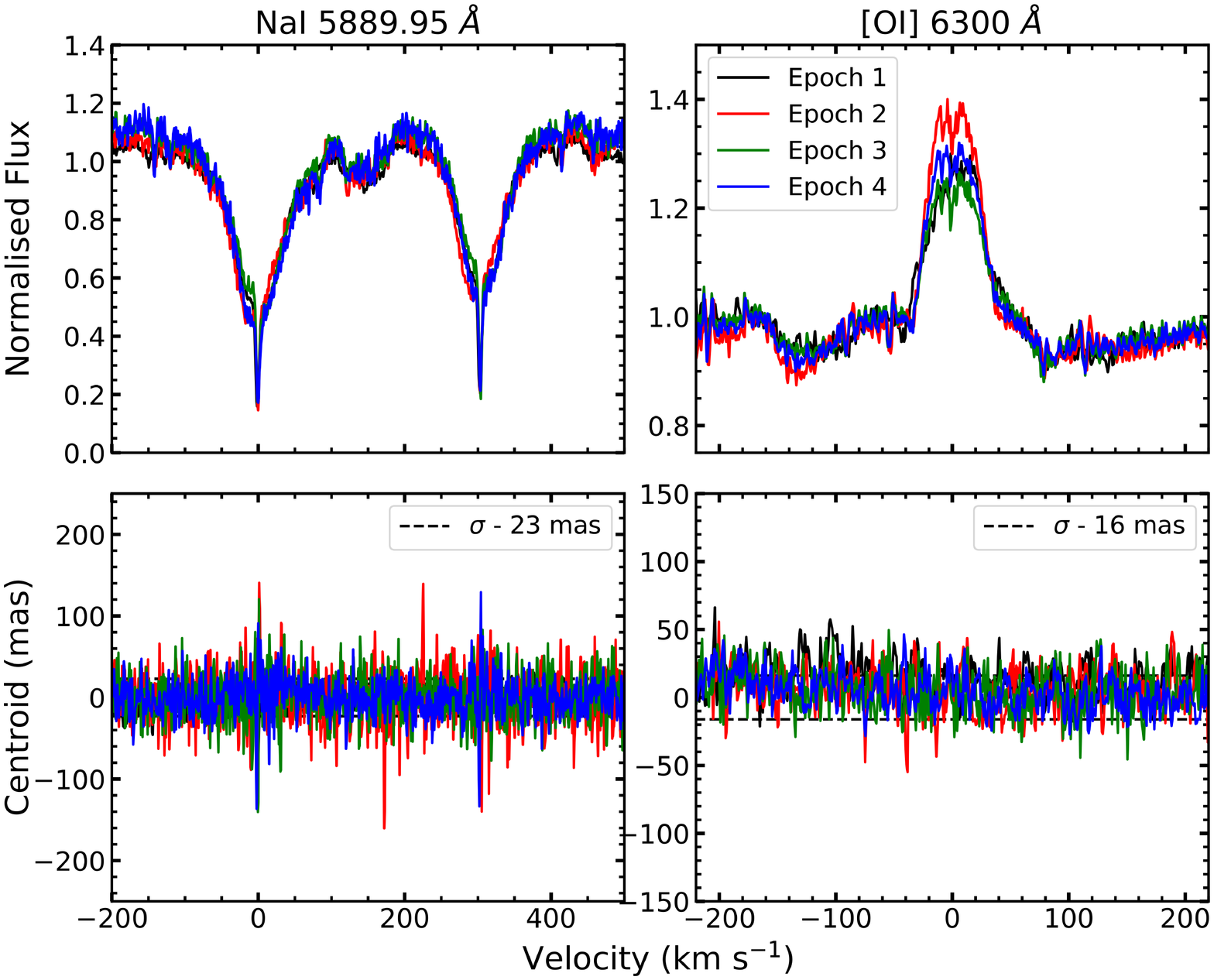}
	\caption{Spectro-astrometric analysis of the Na D and [OI]$\lambda$ 6300 lines.}
	\label{fig:OI_Plot}
\end{figure}

In Figure \ref{fig:2d_results} the spectro-astrometric analysis of the median combined H$\alpha$ spectra for each epoch is presented. No spectro-astrometric signal was detected in the individual spectra making up the four epochs or in the median combined spectra.  The spectra shown here were not binned before being analysed and the average 1-sigma value of the centroid is $\sim$ 4~mas.  SA was also applied to both the [OI]$\lambda$ 6300 lines and the Na D lines but also no signal was detected (see Figure \ref{fig:OI_Plot}). For the [OI]$\lambda$ 6300 line the non-detection rules out the origin in a jet but not in a wind. Thus, this result supports the use of the [OI]$\lambda$ 6300 line as a measure of $\dot{M}_{acc}$ (see Section 4). The signal from a planetary companion is attenuated by the strength of the H$\alpha$ emission from T Cha and by the continuum emission. This would mean that the best chance of detecting the planet is in E2 where both a significant part of the stellar H$\alpha$ is obscured and the continuum emission is minimised. To further investigate E2 the continuum emission was subtracted however no signal was detected. Several different binnings were also tested to increase the S/N and thus spectro-astrometric accuracy. Again no signatures were detected. 

The size of the binning which is appropriate is set by the width in velocity of the emission line region which we want to detect and the spectral resolution of the instrument. For TTS the accretion cut-off for the width of the H$\alpha$ line is generally set at 270~km~s$^{-1}$ with a width of 200~km~s$^{-1}$ adopted for substellar objects \citep{Jay03}. If we assume the same accretion criterion for planetary mass objects and brown dwarfs a bin of 40 would increase the spectro-astrometric accuracy to $\sim$ 0.5~mas while providing a sampling of 5 points across a H$\alpha$ line of width 200~km~s$^{-1}$. The velocity resolution of the UVES spectra is $\sim$~1~km~s$^{-1}$. It is expected that the strongest spectro-astrometric signal would be at the peak of the H$\alpha$ emission line. In Section 5 it is discussed whether this accuracy would be enough to detect a signal from a possible planet. The non-detection of the companion in the H$\alpha$ line would also support the argument that T Cha b does not exist \citep{Sallum2015b}.

\begin{figure*}
\centering
	\includegraphics[width=18cm]{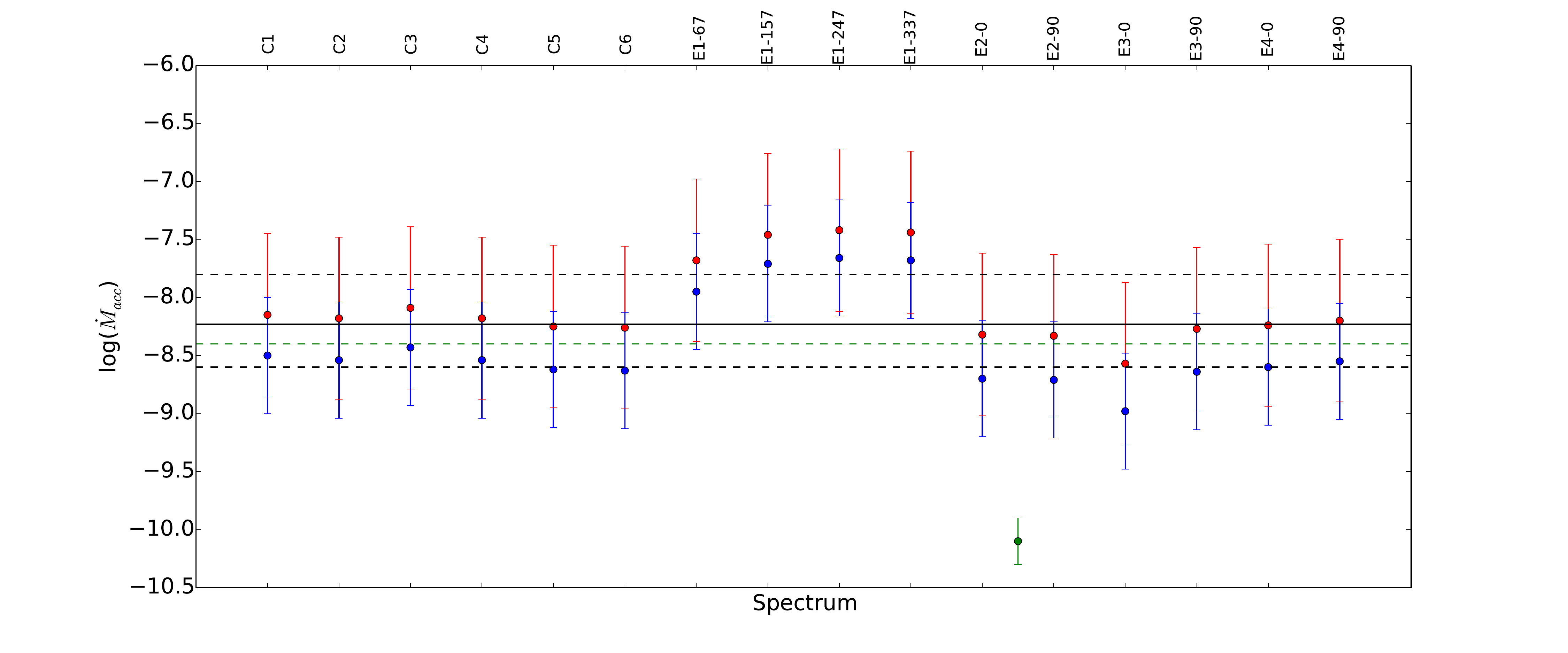}
	\caption{$\dot{M}_{acc}$ rates {\bf (in units of \Msun / yr)} for T Cha calculated using the relationships of Nisini et al. (2018). The blue points correspond to the values calculated using the log(L$_{[OI]LVC}$), log(L$_{acc}$) relationship and the red points to the log(L$_{[OI]LVC}$), log($\dot{M}_{acc}$) relationship. The 10 UVES epochs are included in the calculation as well as the 6 epochs given in Table 3 of Schisano et al. (2009) and taken from Covino et al. (1996), labelled here C1 to C6. The black line is the mean value with the 1-$\sigma$ value shown by the dashed black lines. The green dashed line is the Schisano et al. (2009) result and the green point is the value calculated using the luminosity of the part of the H$\alpha$ line in emission in E2.}
	\label{fig:MassAccretion}
\end{figure*}

\section{Mass Accretion Rate of T Cha}
T Cha is known to lack both detectable veiling and UV excess which is indicative of a low mass accretion rate \cite{Sch09}. However, the high inclination angle of the T Cha disk could mean that a significant amount of the emission from the accretion zone could be blocked by the circumstellar disk giving the impression of a low $\dot{M}_{acc}$. The effect of a high inclination angle disk on the accretion properties has been reported for other objects \citep{Whe14b}. Also \cite{Sch09} note that the non-detection of veiling in T Cha may also be caused by the relatively high continuum flux emitted by its G8 photosphere. \cite{Sch09} also estimate the mass accretion rate for T Cha at log($\dot{M}_{acc}$) = -8.4~\Msun /yr using the 10$\%$ width of the H$\alpha$ line. However, there were several uncertainties associated with this estimate, including the difficulty in defining the continuum level, the fact the relationship between $\dot{M}_{acc}$ and the 10$\%$ width of the H$\alpha$ line (that was derived for the substellar mass regime) shows a larger dispersion at higher masses and finally that the relationship could only be applied where Gaussian decomposition of the line profile into an emission component with overlapping absorptions was possible. As the H$\alpha$ line is primarily detected in absorption here it is not possible to use the H$\alpha$ 10$\%$ width or line luminosity to estimate $\dot{M}_{acc}$, although the latter is attempted here for E2.

An alternative is to use the [OI]$\lambda$ 6300 line which is considered an indirect tracer of $\dot{M}_{acc}$ in TTSs \citep{Whe14b,Nisini18}. To calculate $\dot{M}_{acc}$ from the [OI]$\lambda$ 6300 \AA{} line the new relationships of \cite{Nisini18} giving the luminosity of the [OI]$\lambda$ 6300 LVC in terms of both log(L$_{acc}$) (Figure \ref{fig:MassAccretion} blue points) and log(M$_{acc}$) (Figure \ref{fig:MassAccretion} red points) were used. Also included in this calculation are the \cite{Cov96} [OI]$\lambda$ 6300 line luminosities reported in Table 3 of \cite{Sch09} and given here for completeness in Table \ref{Table2}.  The mean value of log($\dot{M}_{acc}$) calculated from the \cite{Nisini18} relationships (denoted by the blue markers in Fig \ref{fig:MassAccretion}) is -8.3 $\pm$ 0.4~\Msun / yr and this is marked in Figure \ref{fig:MassAccretion} by the black line. The standard deviation in the results is delineated by the black dashed lines. The green dashed line is the average $\dot{M}_{acc}$ value from \cite{Sch09} and the green point is the value calculated using the luminosity of the part of the H$\alpha$ line in emission in E2. The mass accretion rate derived using the H$\alpha$ flux in E2 is -10.1 $\pm$ 0.2~\Msun / yr. For this calculation the relations of \cite{Alcala17} were used. Assuming for the time being that the H$\alpha$ emission line region must originate from outside of the occulting zone, no extinction correction was applied. 

\begin{figure*}
\centering
	\includegraphics[width=15cm]{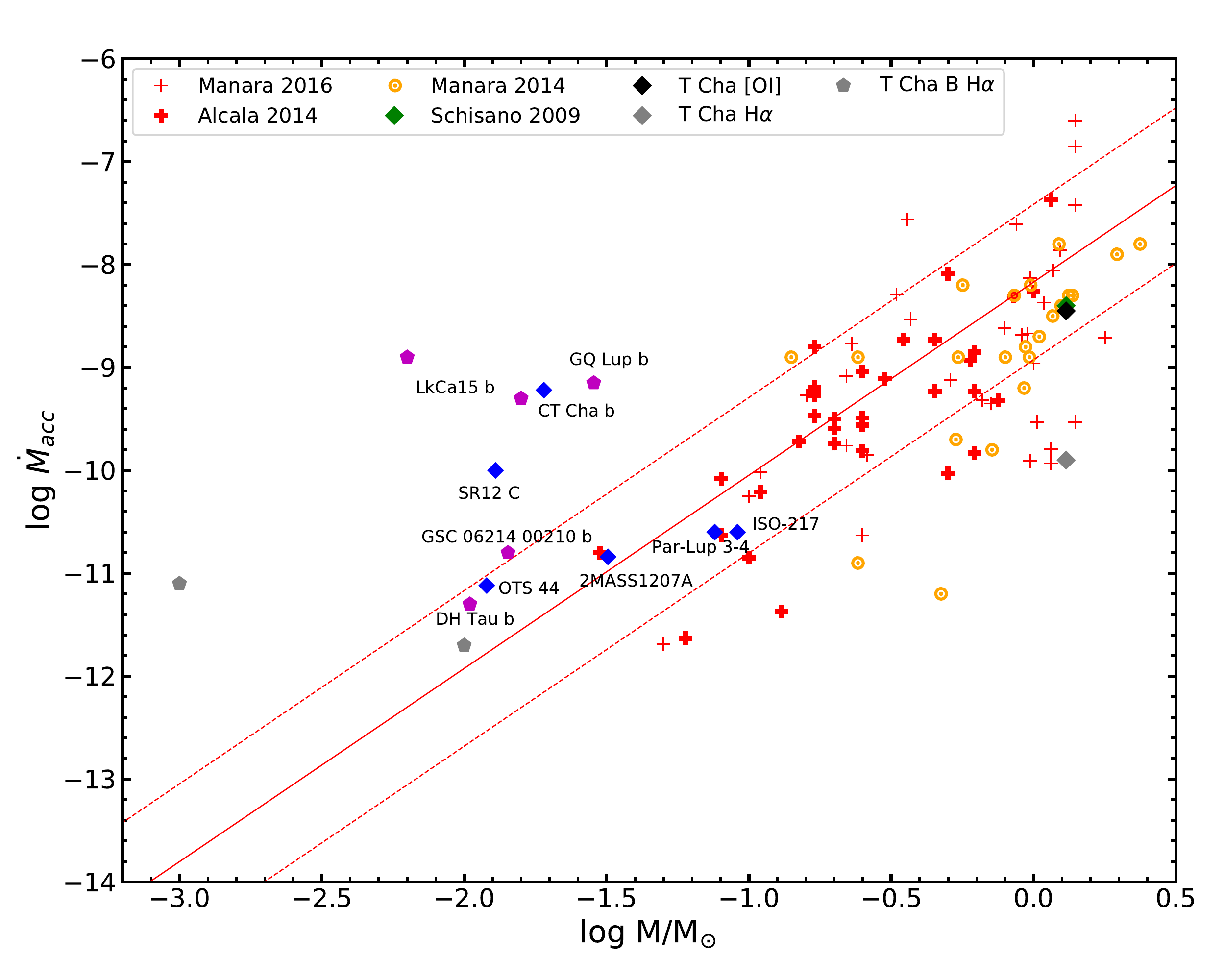}
	\caption{Comparison of the $\dot{M}_{acc}$ values {\bf (in units of \Msun / yr)} obtained for T Cha with accreting objects with masses within the range of low-mass stars to planets. The planets are represented by hexagons and the BDs blue diamonds and they are all marked with their names. The red line represents the linear fit to the results of Alcal{\'a} et al. (2014) and Manara et al. (2016) for a sample of TTSs. It gives the correlation between M and $\dot{M}_{acc}$ for these types of objects. The dashed lines represent the 1-$\sigma$ deviation from the fit. The yellow dots are a sample of TD objects taken from Manara et al. (2014). The values of $\dot{M}_{acc}$ for the TD objects generally lie within the range of the TTS values. For the individual BDs and planets the results are taken from the following studies. Lk~Ca15~b - Whelan et al. (2015); GQ~Lup~b, GSC~06214 and DH~Tau~b - Zhou et al. (2014); OTS~44 - Joergens et al. (2013); 2MASS1207A - Mohanty et al. (2005); ISO-217 - Whelan et al. 2014b; Par-Lup3-4 - Whelan et al. 2014a; SR~12~C - Santamari{\'a}-Miranda et al. (2018); CT~Cha~B - Wu et al. (2015).}
	\label{fig:MaccvM}
\end{figure*}

In Figure \ref{fig:MaccvM} the relationship between mass and mass accretion rate is plotted for low mass and substellar objects, including a sample of TD objects taken from \cite{Manara14}. T Cha is also shown here. The thick and thin red crosses correspond to the TTSs published in \cite{Manara16} and \cite{Alc14} respectively. The \cite{Manara16} sample are all Cha~I sources. The linear correlation shown is for these objects and the motivation is to see how well the TD objects (especially T Cha), brown dwarfs (BDs) and planetary mass objects follow this correlation. This figure builds on figures published in these papers and on Figure 7 of \cite{Whe15}. From Figure \ref{fig:MaccvM} it can be seen that the values of $\dot{M}_{acc}$ for T Cha calculated from the L$_{[OI]}$, the H$\alpha$ 10$\%$ width and the L$_{H\alpha}$ in E2 are within the range of values reported for TDs objects, although the L$_{H\alpha}$ value is on the low side. Comparing the E2 H$\alpha$ $\dot{M}_{acc}$ with the other values would suggest that a significant portion of the H$\alpha$ emission line region is also obscured during the periods of occultation. It is calculated that the H$\alpha$ line flux measured in E2 would need to increase 30 fold to give log($\dot{M}_{acc}$) = -8.3~\Msun / yr. Studies of other YSOs have shown that the disk can greatly suppress emission from the accretion zone leading to a perceived low $\dot{M}_{acc}$ \citep{Nisini18, Whe14b}. The distance of a few tenths of an au proposed by \cite{Sch09} as the distance at which the occulting clumps around T~Cha orbit would be sufficient to block emission from the accretion zone as well as the star and thus can explain the large dimming of the H$\alpha$ line emission. Accretion is expected to occur on a scale of $\sim$ 0.1~au \citep{Har16}. Therefore, the bulk of the H$\alpha$ emission must come from accretion here with perhaps also a small contribution from a wind. Whereas a greater portion of the [OI]$\lambda$ 6300 line is originating from beyond the region of obscuration making it a better measure of $\dot{M}_{acc}$ in this case \citep{Nisini18}.

\section{Mass accretion rate of T~Cha~b and spectro-astrometric accuracy}

Here the results of Figure \ref{fig:MaccvM} are used to constrain a range of values for $\dot{M}_{acc}$ onto T~Cha~b. Once $\dot{M}_{acc}$ is estimated some limits for the spectro-astrometric accuracy needed to detect T~Cha~b can be inferred from the corresponding H$\alpha$ line luminosities. It is assumed that 10~$M{_{Jup}}$ is a reasonable upper limit for the mass of T~Cha~b. Both LkCa~15~b and DH~Tau~b have suggested masses of $\sim$ 10~${_{Jup}}$ and the large difference in their $\dot{M}_{acc}$ could be explained by their different evolutionary stages. LkCa~15~b is an embedded protoplanet while DH~Tau~b is a wide companion \citep{Whe15, Zhou14}. As the evolutionary stage of T~Cha~b is not clear the upper and lower limits to $\dot{M}_{acc}$ offered by LkCa~15~b and DH~Tau~b in Figure \ref{fig:MaccvM} are taken as a reasonable range of values for T~Cha~b. 

\cite{Whe15} calculated log($\dot{M}_{acc}$) = -9.3 for LkCa~15~b which corresponds to L$_{H\alpha}$ $\sim$ 7~$\times$~10$^{-5}$~\Lsun. \cite{Zhou14} measured log($\dot{M}_{acc}$) = -11, -9.3 and -11.4 for GSC~06214~-00210, GQ~Tau~b and DH~Tau~b respectively, using the luminosity of their H$\alpha$ emission. These values of $\dot{M}_{acc}$ correspond to L$_{H\alpha}$ = 9.3~$\times$~10$^{-6}$~\Lsun, 2~$\times$~10$^{-5}$~\Lsun{} and 6.5~$\times$~10$^{-7}$~\Lsun. Also \cite{Zhou14} give the masses of these three objects as 15~M$_{Jup}$, 28~M$_{Jup}$ and 11~M$_{Jup}$ respectively. Keeping to our assumption of a mass of $\sim$ 10~M$_{Jup}$ for T~Cha~b, a value L$_{H\alpha}$ = 1~$\times$~10$^{-6}$~\Lsun\ would seem reasonable as a lower limit given the results of \cite{Zhou14}. Using the 1-$\sigma$ upper limit to the correlation presented in Figure \ref{fig:MaccvM} would suggest a similar value. It is found from our mass accretion rate study that L$_{H\alpha}$ = 4~$\times$~10$^{-3}$~\Lsun\ would be needed to achieve log($\dot{M}_{acc}$) = -8.3~\Msun / yr for the parent star T~Cha. Therefore the ratio between the H$\alpha$ emission from T~Cha and T~Cha~b can be taken as ranging from 1.8~$\times$~10$^{-2}$ to 2.5~$\times$~10$^{-4}$.

Given that \cite{Hen18} show the gap to extend from 18~au to 28~au one can expect any planet to be located at $\sim$ 23~au or 190~mas at the distance to T~Cha. The corresponding size of the offset that would be measured using SA is this distance multiplied by the ratio of the strength of the H$\alpha$ emission from the planet and star which, following the argument laid out above would be 3.4~mas to 0.05~mas. It should also be noted that calculation outlined here assumes no contribution from the continuum emission from T~Cha and that \cite{Hen18} suggest a possible mass of $\sim$ 1~M$_{Jup}$ for T~Cha~b. Both of these points would mean that the signature of T~ Cha~b would be even smaller. Given that a maximum binning of 40 is recommended (section 3.2) and would return a spectro-astrometric accuracy of $\sim$ 0.5~mas it is concluded that while it should be theoretically possible to detect a 10 M$_{Jup}$ planet it is not possible to use SA to detect T~Cha~b in this dataset.  It is worth noting that the [OI] $\dot{M}_{acc}$ results were not used to estimate an upper limit to the flux of T~Cha~b and then used to estimate a $\dot{M}_{acc}$ as this would assume that protoplanets have winds like TTSs, which is something that has not been proven.

The analysis outlined above is the opposite to what was done for LkCa 15~b in \cite{Whe15}, where the spectro-astrometric accuracy was used to put a limit on L$_{H\alpha}$ and thus $\dot{M}_{acc}$. This was done by assuming that the H$\alpha$ emission comes from the star plus planet and by using the spectro-astrometric accuracy to put an upper limit on what percentage comes from the planet. Once this is known the line luminosity method is used to derive $\dot{M}_{acc}$. As this approach relies on measuring the H$\alpha$ line luminosity this would only be possible here for E2 where there is some H$\alpha$ in emission. It is reasonable to assume the H$\alpha$ emission from the planet to be centrally peaked and therefore the emission profile could overlap with the absorption feature seen in Figure 6 and causing infilling of the feature. As a result it was decided that the amount of H$\alpha$ emission from the star could not be reliably estimated. However, if we follow the LkCa 15~b approach using the H$\alpha$ emission present in E2, L$_{H\alpha}$ $\sim$ 7~$\times$~10$^{-7}$~\Lsun is derived. Using the argument made above this would give a necessary spectro-astrometric accuracy of $<$ 0.1~mas in agreement with the lower end of the range given above. This corresponds to a range of log($\dot{M}_{acc}$) of -11.1~\Msun / yr to -11.7~\Msun / yr for a mass range of 1~M$_{Jup}$ to 10~M$_{Jup}$ (grey hexagons Figure 9).


\section{Conclusions}

The UVES observations of T Cha presented generally agree with the conclusions of the exploratory study of \cite{Sch09}. However, by considering four epochs of data, carrying out a spectro-astrometric analysis and by making a more accurate study of the mass accretion rate early results can be expanded on and the following conclusions made.

\begin{enumerate}

\item{The spectro-astrometric analysis shows that the likely origin for the [OI]$\lambda$ 6300 line is a wind. Thus, it is possible that part of the emission due to this wind could be obscured by the occulting clumps in the disk of T Cha. Indeed, this could be the explanation for the differences noted between the E1 [OI]$\lambda$ 6300 line profiles and the E2-E4 profiles. It is concluded that the [OI]$\lambda$ 6300 is a better tracer of the mass accretion rate onto T Cha as results derived from L$_{[OI]}$ give values within the range calculated for other TD objects.}

\item{A comparison between $\dot{M}_{acc}$ calculated from L$_{[OI]}$ and L$_{H\alpha}$ shows that a significant proportion of the H$\alpha$ emission is suppressed during the periods of obscuration. This result points to the bulk of the H$\alpha$ emission originating from the accretion shocks and the scale over which the obscuration occurs is in agreement with magnetospheric accretion theory. Therefore, it is concluded that T Cha is not an intrinsically weak accretor. It can be argued that the use of the H$\alpha$ 10$\%$ width as a measure of $\dot{M}_{acc}$ gives a value more compatible with the [OI]$\lambda$~6300 line and the true $\dot{M}_{acc}$ rate as the H$\alpha$ line luminosity and not necessarily the line width is affected by the orbiting clumpy material.}

\item{By making a reasonable estimate of $\dot{M}_{acc}$ for T~Cha~b and thus the luminosity of its H$\alpha$ emission it can be shown that the spectro-astrometric accuracy achieved here is not sufficient to make a detection. This is true even if a large binning is considered. Therefore, the spectro-astrometric results cannot rule out the existence of T~Cha~b. Overall, it is concluded that the unique properties of T~Cha i.e. the strong G8 photosphere, the variable continuum emission and the variable H$\alpha$ emission make it more complicated to apply SA to this source. In particular, the fact that the line is often in absorption means that is is difficult to quantify the contribution from the stellar continuum and to remove it.}
\end{enumerate}

To date, two TDs objects have been investigated using SA to determine if this technique can be used to detect a planetary companion. In \cite{Whe15} no detection of Lk~Ca15~b was made and this planet was subsequently observed using differential imaging with the H$\alpha$ emission from Lk~Ca15~b having a line luminosity of 6 $\times$ 10$^{-5}$~\Lsun\ \citep{Sallum15}. In \cite{Whe15} an upper limit of 1$\times$ 10$^{-13}$ erg/s/cm$^{2}$ was estimated for the H$\alpha$ flux of LKca15~b. This corresponds to a luminosity of ~7 $\times$ 10$^{-5}$~\Lsun, which is comparable to the flux detected by \cite{Sallum15}. \cite{Manara14} give log($\dot{M}_{acc}$) =-8.4~\Msun / yr for Lk~Ca15 which corresponds to L$_{H\alpha}$ = 3 $\times$ 10$^{-3}$. Our results for T~Cha and estimates for T~Cha~b are in agreement with these results for Lk~Ca15 and Lk~Ca15~b. \cite{Sallum15} detect Lk~Ca15~b at a distance of 93 $\pm$ 8~mas from its parent star. Considering the line luminosities and this distance would mean that a spectro-astrometric accuracy of $<$ 0.2~mas would have been necessary in \cite{Whe15} to detect this planet. This is much better than what was achieved. As is the case here with T~Cha the analysis of Lk~Ca15~b was also complicated by problems with spectro-astrometric artefacts. From the work described in this paper and the Lk~Ca15 study it can be concluded that while straight-forward in theory it can be difficult to apply SA to the problem of routinely detecting planetary companions to TD objects, due to the sub-milliarcsecond accuracies required. However, as was shown for LkCa~15, SA is useful as it provides limits to the flux of the H$\alpha$ emission from accreting planetary companions. This is important as H$\alpha$ measurements of protoplanets are rare and direct H$\alpha$ imaging observations with the extreme AO necessary are only possible with bright stars.










\end{document}